\documentstyle[12pt]{article}

\begin{document}
\begin{flushright}
OCHA-PP-152\\
KEK-TH-686\\
May (2000)
\end{flushright}

\baselineskip 0.8cm

\begin{center}

{\Large\bf
Chamseddine, Fr$\ddot{o}$hlich, Grandjean Metric\\
\bigskip
and Localized 
Higgs Coupling\\
\bigskip 
on Noncommutative Geometry
}\\

\vspace{1cm}

{\sc R.Kuriki} and {\sc A.Sugamoto}$^\sharp$\\
\medskip
{\it Laboratory for Particle and Nuclear Physics,\\
High Energy Accelerator Research Organization (KEK),\\
Tsukuba, Ibaraki 305-0801, Japan\\ \smallskip $\sharp$
Department of Physics,
Ochanomizu University\\
1-1 Otsuka 2, Bunkyo-Ku Tokyo
112-0012, Japan}

\end{center}

\bigskip 


\bigskip

\thispagestyle{empty}


\begin{abstract}

\bigskip

We propose a geometrized Higgs mechanism based on the
gravitational sector in the Connes-Lott formulation of the standard model,
which has been constructed by Chamseddine, Fr$\ddot{o}$hlic and Grandjean.
The point of our idea is that Higgs-like couplings
depend on the local coordinates of the four-dimensional continuum, $M_4$.
The localized couplings can be calculated by the Wilson loops
of 
the $U(1)_{EM}$ gauge field and 
the connection, which is defined on
$Z_2\times{M_4}$.
\end{abstract}

\clearpage


\section{Introduction}
\indent

There are some proposals as constructive definitions of string theory.
One of them is called ``M-theory''
which is limited to eleven-dimensional supergravity at low energy,
and that corresponds to the strong coupling limit of the 
type-IIA superstring \cite{BFSS}.
More concretely, M-theory is described by the IIA matrix model, 
which governs the dynamics of D0-branes.
One of the others is the IIB matrix model, which has been proposed
by Ishibashi, Kawai, Kitazawa and Tsuchiya \cite{IKKT}.
Recently,
the authors of \cite{AIIKKT} have shown that
the twisted reduced model can be interpreted as noncommutative Yang-Mills theory.
They have obtained noncommutative Yang-Mills theory with D-brane
backgrounds in the IIB matrix model. 
Furthermore, some of the same authors have calculated 
the Wilson loop in noncommutative Yang-Mills, and have found
an open string-like object \cite{IIKK}, whose length is $|C^{\mu\nu}k_\nu|$.
$C^{\mu\nu}$ delineates the noncommutative scale. 
In \cite{IKK}, a non-local basis,
called bi-local basis, is introduced 
to provide a simple description for high-momentum $|k^\mu|>\lambda$, where
$\lambda$ is the spacing quanta and $k^\mu$ is the eigenvalue of adjoint
$\hat{P}^\mu$.                            
As can be recognized above, 
some new mathematical methods are required to more exactly investigate
superstring theory.
One of new mathematics is noncommutative geometry, 
which was introduced to physicists 
by a mathematician, Alain Connes \cite{C}.

There are many papers in which authors apply noncommutative geometry 
to rebuilding ordinary physics. The initiative is as follows. 
A trial for constructing the standard model, 
$U(1)\times{SU(2)}\times{SU(3)}$,
has been partially successed in 
the framework of noncommutative geometry \cite{CL}. 
In \cite{CL} the base manifold is $E_4(M)\times{Z}_2$ and 
the algebra is ${\cal A}$=${C}^{\infty}(M)\otimes\{M_2(C)\oplus{M}_1(C)\}$.
The acting Hilbert space is given by that of spinors of 
$\left(\begin{array}{c}
l\{{\rm doublet}\}\\
e\{{\rm singlet}\}
\end{array}
\right)$.
To introduce an $SU(3)$ gauge group for the quarks color,
the authors have taken notice
of the bimodule structure relating two algebras
$({\cal A},{\cal B})$, where ${\cal A}$=${\rm algebra~of~quaternions}
\oplus{C}$ and ${\cal B}$=$C\oplus{M}_3(C)$. Finally,
the Higgs field appears as a gauge field between two points 
represented by $Z_2$.  From the unification of 
the standard model and gravitational theory,
this work is very meaningful, because it gives a geometrical interpretation
of the standard model.  

However, the Connes-Lott formulation \cite{CFG}
could not be beyond the standard model,
as they have already described in the literature.
For example, the generation matrix is given by
$C^{N_G}$; namely, the authors could not introduce the inside structure
of the generation.
In \cite{C} A.Connes represents ultimate noncommutative space including 
the color symmetry $SU(3)$
as the 4-dimensional continuum times the finite space `F'.
What is the quantum group of the finite space `F' ?
This is one of problems which
he discusses on the last page of \cite{C},
which one should solve when
trying to make up a beyond standard model on noncommutative
geometry.
We have already recognized that
we could not excel A.Connes in mathematical abilities.
Hence, we attack the open problem from the opposite side to
mathematicians.

We expect that noncommutative geometry may better extract
the potential abilities of the existing physical models, 
for instance the standard model,
though there are still a few papers which have suggested that 
noncommutative geometry can expose a new thing in ordinary things.
In the future we would like to 
determine the masses of quarks based on geometric quantities
on some noncommutative algebraic structure.
Our purpose is to propose a gravitational model coupled with
an $SU(2)$ doublet complex Higgs-like field.
In this paper we artificially
elevate the ordinary Higgs couplings to some components of
one-form basis for our noncommutative space.
We then impose the unitarity condition and the torsion-less condition
for the basis of our noncommutative differential geometry as the authors 
of \cite{CFG}\cite{CFF} have performed.
Finally, in virtue of imposing these conditions, 
we will find that our Higgs-like couplings  
are represented by the Wilson loops
of the spin connections of the noncommutative space and $U(1)_{EM}$
gauge fields.
These gauge fields are expressed by the linear combinations of
the $U(1)_Y$ and $SU(2)_W$ gauge fields.
In section two we briefly review \cite{CFG}.
In section three we will introduce our vielbein 
while comparing it to Chamseddine, Fr$\ddot{o}$lich, Grandjean's
metric. Our discussion is given in the last section.
The unitarity conditions and the components of torsion  
are expressed in the appendices. 


\section{Review of Chamseddine, Fr$\ddot{o}$hlich, Grandjean metric}

In this section
we introduce \cite{CFG} in order to describe the underlying 
geometric structure of our proposal.
The authors have studied the gravitational action of the noncommutative 
geometry underlying the above Connes-Lott construction 
of the standard model \cite{CFG}.
Ultimately, they have found the Einstein-Hilbert
action in the leptonic sector and one in the quark sector.
In these actions they have shown that the distance function between the 
two points is determined by a real scalar field, $\sigma$,
whose vacuum
expectation value sets the weak scale.

They set up the noncomutative differential geometry on 
two copies of the four-dimensional continuum, $M_4$,
defined by
\begin{eqnarray}
M_4\times{Z}_2.
\label{MZ}
\end{eqnarray}
They choose the algebra ${\cal A}$ describing the above noncommutative space
as 
\begin{eqnarray}
{\cal A}=({\cal A}_1\oplus{\cal A}_2)\otimes{C}^\infty(M_4),
\end{eqnarray}
where ${\cal A}_1=M_2(C)$
and ${\cal A}_2=C$. 
They firstly consider the 
leptonic part and the Higgs sector of the standard model. 

$a\in{\cal A}$ is represented by
\begin{eqnarray}
a=
\left(
\begin{array}{cc}
a_1&0\\
0&a_2
\end{array}
\right),
\end{eqnarray}
where $a_1$ (for ${\cal A}_1$),$a_2$ ( for ${\cal A}_2$) 
are the $C^{\infty}$ function on $M_4$. Notice that $a_1$ is a two times two
matrix.

They define the representation space for ${\cal A}$ and the Hilbert space
as follows:
\begin{eqnarray}
{\cal H}=L^2(S_1,dv_1)\oplus{L}^2(S_2,dv_2),
\label{Hilbert}
\end{eqnarray}
where 
$S_i=S_0\otimes{V}_i$. $S_0$ is the bundle of Dirac spinors on $M_4$
and $V_i$ is a representation space for ${\cal A}_i$. They give 
the representation space as $V_1=C^2, V_2=C$. $dv_i$ is the volume
form corresponding to a Riemannian metric on $M_4$.

They express the Dirac operator on the noncommutative space as
\begin{eqnarray}
D=\left(
\begin{array}{cc}
\nabla_1\otimes{\bf 1}_2\otimes{\bf 1}_3&
\gamma^5\otimes{\bf M}_{12}\otimes{\bf k}\\
\gamma^5\otimes{\bf M}^{\ast}_{12}\otimes{\bf k}^\ast&
\nabla_2\otimes{\bf 1}_3
\end{array}
\right),
\end{eqnarray}
where $\nabla_i$ is defined by
\begin{eqnarray}
\nabla_i=e^\mu_{ia}\gamma^a(\partial_\mu+i\omega_{i\mu}).
\end{eqnarray}
$e^\mu_{ia}$ is a vielbein of $M_4$. The index $i=1,2$ distinguishes 
${\cal A}_i$ and $\mu$ represents the vector index on $M_4$; 
$\mu$=0,1,2,3; $a$ is the coordinate of
the tangent bundle $M_4$; $a$=1,2,3,4.
$\omega_{i\mu}$ is the spin connection in (\ref{MZ}).
$\{\gamma^a\}^4_{a=1}$ are the anti-hermitian Euclidean Dirac matrices,
with
$\{~\gamma^a~,~\gamma^b~\}=
\gamma^a\gamma^b+\gamma^b\gamma^a=-2\delta^{ab}$,
$
\gamma^5=\gamma^1\gamma^2\gamma^3\gamma^4$,
$\gamma^5=(\gamma^5)^\ast$,
and
$-\gamma^a=(\gamma^a)^\ast$.
$\gamma^\ast$ means the hermitian conjugate.
We follow their notation in the section three.
The components of the spin connection are chosen based on 
the Cartan structure equation of (\ref{MZ}) as the Riemannian geometry.   
$k$ is a $3\times{3}$ family mixing matrix.
In the leptonic sector
the detailed structure inside of $k$ is not necessary.
$M_{12}$, called a doublet, is written as 
$
M_{12}=\left(
\begin{array}{c}
\alpha(x)\\
\beta(x)
\end{array}
\right).
$
Functions $\alpha(x)$ and $\beta(x)$ 
are restricted by the requirements the consistency of 
$\Omega^2_D({\cal A})$, which is isomorphism to 
$\pi(\Omega^2({\cal A}))/{\rm Aux}^2$, where
\begin{eqnarray}
{\rm Aux}^2=\{
\sum_i\left[D,a_i\right]\left[D,b_i\right]=0:
\sum_i{a_i}\left[D,b_i\right]=0\}.
\label{aux1}
\end{eqnarray}
In other words, we should pull the following elements $a_i,b_i$
from ${\cal A}$:
\begin{eqnarray}
{\rm If}\quad \rho&=&\sum_i{a_i}d{b_i}~\in~{\rm ker}\pi
\qquad \Longleftrightarrow \pi(\rho)=0,
\nonumber\\
{\rm then}\quad \pi(d\rho)&=&\sum_i\left[D,a_i\right]
\left[D,b_i\right]=0.
\label{aux2} 
\end{eqnarray}
The authors find three possibilities which can satisfy Eq.(\ref{aux2}),
and in detail compute one of these in \cite{CFG},
\begin{eqnarray}
M_{12}=\exp(-\sigma)\left(
\begin{array}{c}
0\\
1
\end{array}
\right),
\end{eqnarray}
where $\sigma(x)$ is a real scaler field.

They introduce a system of generators of  
$\tilde{\Omega}^1_D({\cal A})$, $\{E^A\}$,
which are suitable for the Hilbert space (\ref{Hilbert}) 
and the representation space defined by themselves:
\begin{eqnarray}
E^a&=&\gamma^a\left(
\begin{array}{cc}
{\bf 1}_2&0\\
0&1
\end{array}
\right),\quad a=1,2,3,4,
\label{Ea}
\\
E^r&=&\gamma^5
\left(
\begin{array}{cc}
{\bf 0}_2&ke_r\\
-k^{\ast}{e}^{\top}_r&0
\end{array}
\right),\quad r=5,6,
\label{Er}
\end{eqnarray}
where
$
e_5=\left(\begin{array}{c}
1\\
0
\end{array}
\right)$,
$e_6=\left(\begin{array}{c}
0\\
1
\end{array}
\right),
$
and $\top$ means the transposed matrix.
\footnote{The authors of \cite{CFG} define
$\tilde{\Omega}({\cal A}):=\tilde{\pi}(\Omega^.({\cal A}))$,
$\tilde{\Omega}^n_D({\cal A}):=\tilde{\pi}(\Omega^n({\cal A}))/
\tilde{\pi}(dJ_{n-1})$. Here $J_n$ is the intersection of the kernel 
which is given by Eq.(\ref{aux2}). $\tilde{\pi}$ is a *-representation 
of $\Omega^*({\cal A})$.}

They list up the conditions of vanishing torsion $T^A,A=1,\ldots,6$,
where $T^A=dE^A+\Omega^A{}_BE^B$ and 
the unitarity conditions on (\ref{Ea}) and (\ref{Er}).
The components $\Omega^A{}_B$ which are defined by
\begin{eqnarray}
\nabla{E}^A=-\Omega^A{}_B\otimes{E}^B,
\label{connection}
\end{eqnarray}
are the connection 
coefficients on $\tilde{\Omega}^1_D({\cal A})$ \cite{CFG}.
After imposing these conditions, they declare that
the field $\sigma$ becomes non-dynamical.
They then weaken the condition of vanishing torsion as 
\begin{eqnarray}
Tr_k{T}^A=0
\label{weak}
\end{eqnarray}
where $Tr_k$ is the trace over the family mixing matrix. 
As the result of Eq.(\ref{weak}), the $\sigma$ 
field behaves as a dynamical field
in the Hilbert-Einstein action of the leptonic sector.

They also construct the gravitational action of the quark sector.
It is the same form of action of the leptonic sector, but with
different coefficients and with dependence on the generation-mixing 
matrices of the quarks masses.
The action is given by
\begin{eqnarray}
\int{d^4x}\left[
-\frac{1}{2}(3c_l+4c_q){R}+\alpha(\nabla_a\sigma)^2+\beta{e}^{-2\sigma}
\right],
\label{gravitational}
\end{eqnarray}
where
$c_l,c_q$ are arbitrary constants and 
$\alpha$,$\beta$ depends on the structure of the generation matrices 
of the quark masses and $c_l,c_q$ \cite{CFG}.
However, as the authors of \cite{CFG} described, 
since the gravitational action is non-renormalizable 
and nobody understands the quantum noncommutative geometry,
these coefficients do not have any physical significance. 
$R$ is the scalar curvature in $M_4\times{Z_2}$.

In string theory we have known a theory whose action is almost
the same form as (\ref{gravitational}). 
It is the Liouville field theory, where the action contains
an exponential tachyon field, $\exp{(\alpha_{-}x^1)}$. Here,
$x^1$ is one-direction of the target space, and the effective string
coupling is diverging at large $x^1$.
$\alpha_{-}$ is given by 
$(\frac{26-D}{6\alpha'})^{1/2}-(\frac{2-D}{6\alpha'})^{1/2}$.
For $D>2$ this is complex, hence this term oscillates, 
but for $D\leq{2}$, $\alpha_{-}>0$,
we have a real exponential.
\footnote{In the discussion of the Liouville field theory, we have
referred to the explanation on pp.323-327 of \cite{Po}.} 
The difference between (\ref{gravitational})
and the Liouville action is that
the latter contains the term $R\times{x^1}$. 
The difference between the sign of the exponential interactions
can be eliminated by redefining of the field $\sigma$ to $-\sigma$.
What we want to 
stress here is that in string theory 
we may express the degree of freedom, $x^1$,
as a differential operator for some discrete space, 
since the authors of \cite{CFG} 
can gain (\ref{gravitational}), 
which is similar to the Liouville action, from 
the gravitational theory on $M_4\times{Z_2}$.

\section{Our proposal- geometrized Higgs mechanism} 
\indent

An outline of our idea is that
we would like to finally generalize 
an ordinary Higgs mechanism of the standard model
to a gravitational theory on a noncommutative geometry.
We attempt to raise the Higgs coupling from a constant
to one of components of vielbein on noncommutative geometry,
$M_4\times{Z_2}$.
Furthermore, we make our Higgs couplings
\footnote{We call our Higgs couplings Higgs-like couplings.}
 depend on the local coordinates
of $M_4$.
In the second subsection 
we impose the torsion-less condition and the unitarity 
condition on the vielbein and the spin connection of $M_4\times{Z_2}$,
following \cite{CFG}.
As a result, we find that the Higgs-like couplings 
can be calculated by
the Wilson loops of the spin connection and the $U(1)_{EM}$
gauge.

\subsection{Dynamical vielbein and localized Higgs couplings}

We choose the algebra ${\cal A}$, defining the noncommutative space
underlying our model as
\begin{eqnarray}
{\cal A}=({\cal A}_1\oplus{\cal A}_2\oplus{\cal A}_3)
\otimes{C}^{\infty}(M_4),
\label{ouralgebra}
\end{eqnarray}
where $M_4$ is a smooth, compact, four-dimensional Riemannian spin
manifold; ${\cal A}_1$ is  
the algebra of complex $2\times{2}$ matrices.
${\cal A}_2$ and ${\cal A}_3$ are $C$.
Elements, $a$, of ${\cal A}$ are written as
\begin{eqnarray}
a=
\left(\begin{array}{ccc}
a_1&0&0\\
0&a_2&0\\
0&0&a_3
\end{array}
\right),
\end{eqnarray}
where $a_i$ is a $C^\infty$-function on $M_4$ with values
in ${\cal A}_i$, $i=1,2,3$.

The Hilbert space is defined as the spinors of the form 
$L=\left(\begin{array}{c}u_L\\d_L\\u_R\\d_R\end{array}\right)$,
where $R,L$ respectively mean the two kinds of chiralities 
which are defined by the four-dimensional $\gamma^5$.
We want to regard $u_R,d_R$ as an $SU(2)$ singlet and 
$u_L,d_L$ as an $SU(2)$ doublet. Here, we image the Hilbert space
of one-family quarks.

We introduce the following basis as
\begin{eqnarray}
{\cal E}^a&=&\gamma^a
\left(
\begin{array}{cccc}
1&0&0&0\\
0&1&0&0\\
0&0&1&0\\
0&0&0&1
\end{array}
\right),
\label{cEa}
\\
{\cal E}^{\bar{r}}&=&\gamma^5
\left(
\begin{array}{cc}
{\bf 0}&\bar{\cal A}{e}_{\bar{r}}\\
-e^\ast_{\bar{r}}\bar{\cal A}&{\bf 0}
\end{array}
\right),
\label{Ebar}
\end{eqnarray}
where 
\begin{eqnarray}
\bar{\cal A}&=&
\left(
\begin{array}{cc}
0&\frac{v}{\surd{2}}\\
\frac{v}{\surd{2}}&0
\end{array}
\right),
\\
e_u=e_{\bar{u}}&=&
\left(
\begin{array}{cc}
0&0\\
1&0
\end{array}
\right),\quad
e_d=e_{\bar{d}}=
\left(
\begin{array}{cc}
0&1\\
0&0
\end{array}
\right),
\end{eqnarray}
\begin{eqnarray}
e^\ast_u=e^\ast_{\bar{u}}=e_d=e_{\bar{d}},\quad
e^\ast_d=e^\ast_{\bar{d}}=e_u=e_{\bar{u}}.
\end{eqnarray}
$a=1,\ldots,4$ and $\bar{r}=\bar{u},\bar{d}$. We call this basis flat basis.
(3.25) and (3.26) in \cite{CFG}
inspire us (\ref{Ea}) and (\ref{Ebar}).
$v$ is a constant.
We would like to first treat (\ref{cEa}) and (\ref{Ebar})
as a system of generators of 
the one-form $\tilde{\Omega}^1({\cal A})$
though (\ref{cEa}) 
and (\ref{Ebar}) contain auxiliary parts as we will explain in the section $4$.

Let us introduce a local basis $e^a_\mu(x)$, $\mu=1,\ldots,4$ 
of orthonormal tangent vectors to
$M_4$ for each $a$. $a$ is the local Lorentz index 
on the flat tangent plane.
In curved $M_4$ we introduce gamma matrices by 
$\gamma^\mu\equiv{e}^\mu_a\gamma^a$ and define a curved base 
which we would like to regard 
as one-form
of $\tilde{\Omega}^1({\cal A})$,
\begin{eqnarray}
{\cal E}^\mu=\gamma^\mu
\left(
\begin{array}{cccc}
1&0&0&0\\
0&1&0&0\\
0&0&1&0\\
0&0&0&1
\end{array}
\right).
\label{curved}
\end{eqnarray}
Moreover, let us consider remaining components  for the curved base. 
These contain Higgs-like couplings
$f(x)$, $\tilde{f}(x)$, which depend on the local
coordinates of $M_4$ as
\begin{eqnarray}
{\cal E}^r=\gamma^5
\left(
\begin{array}{cc}
{\bf 0}&{\cal A}^\ast{e}_r\\
-e^\ast_r{\cal A}&0
\end{array}
\right),
\label{higgs}
\end{eqnarray}
where
\begin{eqnarray}
{\cal A}&=&
\left(
\begin{array}{cc}
0&{f(x)\frac{v}{\surd 2}}\\
{\tilde{f}(x)\frac{v}{\surd 2}}&0
\end{array}
\right),\quad
{\cal A}^\ast=
\left(
\begin{array}{cc}
0&{\tilde{f}^\ast(x)\frac{v}{\surd 2}}\\
{f^\ast(x)\frac{v}{\surd 2}}&0
\end{array}
\right).
\end{eqnarray}
We express the dual basis, following 
(3.29) of \cite{CFG}, 
\begin{eqnarray}
{\bf \omega}=
\left(
\begin{array}{cccc}
{\gamma^\mu\omega_{1\mu,ij}(x)}&~&{{\omega'}_2(x)}&0\\
~&~&0&{\omega_2(x)}\\
{{\omega'}_1(x)}&0&{\gamma^\mu\omega_{2\mu,ij}(x),}&~\\
0&{\omega_1(x)}&~&~
\end{array}
\right),
\end{eqnarray}
where $\omega_{1\mu,ij}(x)$ and $\omega_{2\mu,ij}(x)$ are two-times-two
matrices, $i,j=1,2$.

Following (3.28) and (3.29) in \cite{CFG},
we write down the following connection coefficients
of (\ref{curved}) and (\ref{higgs}), 
which have been defined in (\ref{connection}), as  
%
%
\begin{eqnarray}
\Omega^A{}_B=\left(
\begin{array}{cc}
{\gamma^\mu\omega_{1\mu}{}^A{}_B,{}_{ij}}&
{\cal A}\gamma^5{e}^{-\sigma}
\left(
\begin{array}{cc}
{\omega'}_2{}^A{}_B&0\\
0&\omega_2{}^A{}_B
\end{array}
\right)\\
{
\left(
\begin{array}{cc}
{\omega'}_1{}^A{}_B&0\\
0&\omega_1{}^A{}_B
\end{array}
\right)
\gamma^5e^{-\sigma}{\cal A}^\ast
}
&
{\gamma^\mu\omega_{2\mu}{}^A{}_B,{}_{ij}}
\end{array}
\right),
\end{eqnarray}
\begin{eqnarray}
\Omega^A{}_B{}^\ast
=
\left(
\begin{array}{cc}
{-\gamma^\mu
\omega_{1\mu}{}^A{}_B{}^{\top}
}
&{{\cal A}\gamma^5e^{-\sigma}
\left(
\begin{array}{cc}
{\omega'}_1{}^A{}_B&0\\
0&\omega_1{}^A{}_B
\end{array}
\right)
}
\\
{
\left(
\begin{array}{cc}
{{\omega'}_2{}^A{}_B}&0\\
0&{\omega_2{}^A{}_B}
\end{array}
\right)
\gamma^5e^{-\sigma}{\cal A}^\ast
}&
{-\gamma^\mu\omega_{2\mu}{}^A{}_B{}^{\top}}
\end{array}
\right),
\end{eqnarray}
where $A,B,C=1,2,3,4,5,6$ and
$5,6$ respectively, refers to $\bar{u},\bar{d}$.

We define 
the Dirac operator on the curved space, which is described by 
(\ref{curved}) and (\ref{higgs}), as
\begin{eqnarray}
\hat{D}=
\left(
\begin{array}{cccc}
i\nabla_R&0&&\gamma^5{\bf M}^\ast\\
0&i\nabla_R&&\\
&&&\\
\gamma^5{\bf M}&&&i\nabla_L
\end{array}
\right),\hspace{3cm}&&
\\
{\bf M}^\ast=
\left(
\begin{array}{c}
{f^\ast(x)(~\phi^{0\ast}~,~\phi^+~)}
\\
{\tilde{f}^\ast(x)(~-\phi^-~,~\phi^0~)}
\end{array}
\right),
~~~
{\bf M}=
\left(
\begin{array}{cc}
{f(x)
\left(\begin{array}{c}
\phi^0\\
\phi^-
\end{array}
\right)
}&
{
\tilde{f}(x)
\left(
\begin{array}{c}
-\phi^+\\
\phi^{0\ast}
\end{array}
\right)
}
\end{array}
\right),
\end{eqnarray}
where $\phi^0$ is the neutral Higgs-like field and 
$\phi^{+\ast}=\phi^-$.
\begin{eqnarray}
i\nabla_R=\gamma^\mu(i\partial_\mu-\frac{g'}{2}YB_\mu),
&&\\
i\nabla_L=\gamma^\mu(i\partial_\mu-\frac{g}{2}\tau_a{A}_{a\mu}
-\frac{g'}{2}YB_\mu),
\end{eqnarray}
where $A_{a\mu}(x)$ is the $SU(2)_W$ gauge field 
and $B_\mu(x)$ is the $U(1)_Y$ gauge field.
$\tau_a$ are the Pauli matrices.
\footnote{
$\tau_1=
\left(\begin{array}{cc}
0&1\\
1&0
\end{array}\right)$,
$\tau_2=
\left(\begin{array}{cc}
0&-i\\
i&0
\end{array}\right)$,
$\tau_3=
\left(
\begin{array}{cc}
1&0\\
0&-1
\end{array}
\right)
$.}

\subsection{On-shell Higgs couplings}
Following \cite{CFG} we impose the unitarity condition and 
the torsion-less condition on these gravitational components
given in the previous subsection $3.1$.
First, we would like to impose the unitarity condition for the components.
There are four possibilities from $\alpha$ to $\delta$:
$\alpha.$
(\ref{Ea}) and (\ref{Ebar}),
$\beta.$
(\ref{Ea}) and (\ref{higgs}),
$\gamma.$
(\ref{curved}) and (\ref{Ebar}),
$\delta.$
(\ref{curved}) and (\ref{higgs}). 
We here treat the second case, $\beta$.
In the curved quark space, $u$, $d$, the generalized Higgs couplings depend
on the local coordinate,
$f=f(x),~~~\tilde{f}=\tilde{f}(x)$.

In order to discuss the suitable Dirac operator, which we will 
calculate the unitarity condition, 
let us remember here the spontaneous symmetry breaking down 
in the standard model.
In $SU(2)_W\times{U}(1)_Y$ non-abelian
gauge theory, 
if one gives the following vacuum expectation value of the Higgs
doublet field:
\begin{eqnarray}
\langle{0}\vert{\Phi(x)}\vert{0}\rangle
=\frac{1}{\surd{2}}
\left(
\begin{array}{c}
0\\
v
\end{array}
\right),
\label{higgsvev}
\end{eqnarray}
then the mass terms of the gauge fields appear as
$
M^2_W{W}^\dagger_\mu{W}^\mu+\frac{1}{2}M^2_Z{Z}_\mu{Z}^\mu
$  
in the Lagrangian. 
Here,
$W_\mu=\frac{1}{\surd2}(A^1_\mu-iA^2_\mu)$ corresponds to the W boson
and
$Z_\mu=\cos\theta_W{A}^3_\mu-\sin\theta_W{B}_\mu$ is the Z boson.  From 
the action one can easily recognize that a linear combination of
$SU(2)_W\times{U}(1)_Y$ gauge fields,
$A_\mu=\sin\theta_W{A}^3_\mu+\cos\theta_W{B}_\mu$, is still a mass-less
gauge field. 

Next, let us recall the leptonic sector
after spontaneous symmetry breaking. 
The left-handed sector is
$SU(2)_W$ doublet and the right-handed sector is
$SU(2)_W$ singlet, because the charged current 
weak interaction of leptonic fields 
is the V-A interaction.
When the vacuum expectation value of the Higgs doublet is 
(\ref{higgsvev}), the masses of leptonic fields, $m_j$, 
are in proportion to
the Higgs coupling as $f_j={\surd2}{m_j}/{v}$. 
The masses of quarks are
in proportion to the Higgs couplings, similar to the leptonic sector.
The only different point between the leptonic sector and
the quarks sector is that the lower quarks of the left handed 
($SU(2)_W$ doublet) are not mass eigenstates.

We apply the following Dirac operator on our Hilbert space 
and we will impose the unitarity condition and the torsion-less condition
on our bases,
\begin{eqnarray}
\hat{D}_{EM}=
\left(
\begin{array}{cc}
{\begin{array}{cc}
{i\partial_{EM}}&0\\
0&{i\partial_{EM}}
\end{array}
}&
{\frac{v}{\surd2}\gamma^5
\left(
\begin{array}{cc}
{f^\ast(x)}&0\\
0&{\tilde{f}^\ast(x)}
\end{array}
\right)
}\\
{\frac{v}{\surd2}\gamma^5
\left(
\begin{array}{cc}
{f(x)}&0\\
0&{\tilde{f}(x)}
\end{array}
\right)
}&
{
\begin{array}{cc}
{i\partial_{EM}}&0\\
0&{i\partial_{EM}}
\end{array}
}
\end{array}
\right),
\end{eqnarray}
where
\begin{eqnarray}
i\partial_{EM}&=&\gamma^\mu(i\partial_\mu-{L}A^{EM}_\mu),
\nonumber
\\
A^{EM}_\mu&=&\sin\theta_WA^3_\mu+\cos\theta_W{B}_\mu.
\end{eqnarray}
$L$ is the $U(1)_{EM}$ charge and $\theta_W$ is the 
Weinberg angle.

We summarize the unitarity condition and the components of torsion in
appendices.  From (\ref{tilf1}) 
and (\ref{tilf2}) in appendix A, we ultimately acquire
\begin{eqnarray}
{\mid\tilde{f}\mid}^2(x)
=C_1
\exp{i\int^xdy^\nu
(2\omega_{1\nu}{}^u{}_u,{}_{11}-LA^{EM}_\nu)}&(y),&
\label{tilf1sol}\\
{\mid\tilde{f}\mid}^2(x)
=C_2
\exp{i\int^xdy^\nu
(2\omega_{2\nu}{}^u{}_u,{}_{11}-LA^{EM}_\nu)}(y).
\label{tilf2sol}
\end{eqnarray}
 From (\ref{f1}) and (\ref{f2}) in the appendix,
we obtain
\begin{eqnarray}
{\mid{f}\mid}^2(x)
=C_3
\exp{i\int^xdy^\nu
(2\omega_{1\nu}{}^d{}_d,{}_{22}-LA^{EM}_\nu)}&(y),&
\label{f1sol}\\
{\mid{f}\mid}^2(x)
=C_4
\exp{i\int^xdy^\nu
(2\omega_{2\nu}{}^d{}_d,{}_{22}-LA^{EM}_\nu})(y).
\label{f2sol}
\end{eqnarray}
In the left-hand side,
${\mid{f}\mid}^2$ and ${\mid\tilde{f}\mid}^2$, are real,
we can actually observe these.
On the other hand, we can
calculate the right-hand sides of (\ref{tilf1sol}),
(\ref{tilf2sol}), (\ref{f1sol}) and (\ref{f2sol})
as the Wilson loops of the spin connections,
which is a kind of gauge field and the $U(1)_{EM}$
gauge field. In ordinary gauge theories and string theory
it is well-known that the Wilson loops are one of the observables.
Therefore, we may expect that these equations are geometrically
meaningful for our noncommutative geometry, whose algebra
is $\{M_2(C)\oplus{C}\oplus{C}\}\otimes{C^\infty(M_4)}$.

In the list of the components of the torsion given by  
the appendix B we find some solvable differential equations:
\begin{eqnarray}
T^d_{42}=-\frac{v}{\surd2}
(i\gamma^\mu\gamma^5\partial_\mu{f}\cdot
-\left[\gamma^5,\gamma^\mu\right]_{-}
f{i}\partial_{EM\mu})
-\frac{v}{\surd2}f\gamma^\mu\gamma^5\omega_{2\mu}{}^d{}_{d,22}=0,
\label{39}
&&
\\
T^d_{24}=\gamma^\mu\gamma^5\frac{v}{\surd2}f^\ast\omega_{1\mu}{}^d{}_{d,22}
+\frac{v}{\surd2}(i\gamma^\mu\gamma^5\partial_\mu{f}^\ast\cdot
-\left[\gamma^5,\gamma^\mu\right]_{-}f^\ast{i}\partial_{EM\mu})=0,
&&
\\
T^u_{13}=\gamma^\mu\gamma^5\frac{v}{\surd2}\omega_{1\mu}{}^u{}_{u,11}
\tilde{f}^\ast
+i\frac{v}{\surd2}\gamma^\mu\gamma^5\partial_\mu\tilde{f}^\ast\cdot
+i\frac{v}{\surd2}\tilde{f}^\ast
\left[\gamma^\mu,\gamma^5\right]_{-}
\partial_{EM\mu}=0,
&&
\\
T^u_{31}=
-i\frac{v}{\surd2}\gamma^\mu\gamma^5
\partial_\mu\tilde{f}\cdot-\frac{v}{\surd2}
\tilde{f}\left[\gamma^\mu,\gamma^5\right]_{-}i\partial_{EM\mu}
-\gamma^\mu\gamma^5\tilde{f}\frac{v}{\surd2}\omega_{2\mu}{}^u{}_{u,11}=0,
&&
\label{torsionless}
\end{eqnarray}
where `$\cdot$' means that the differential operator, $\partial_\mu$,
only acts on the next $f$.

If at a quantum level we imposed the torsion-less 
condition on the following ground state
\begin{eqnarray}
\partial_\mu|0>=0,
\end{eqnarray}
Solutions of Eq.(\ref{39})-Eq.(\ref{torsionless}) would be given by:
\begin{eqnarray}
f^\ast(x)=Q_1:\exp{i\int^xdy^\nu(\hat{\omega}_{1\nu}{}^d{}_{d,22}
-2L\hat{A}^{EM}_\nu)(y)}:,
\label{Q1}
&&\\
f(x)=Q_2:\exp{i\int^xdy^\nu(\hat{\omega}_{2\nu}{}^d{}_{d,22}
-2L\hat{A}^{EM}_\nu)(y)}:,
\label{Q2}
&&\\
\tilde{f}^\ast(x)=Q_3:\exp{i\int^xdy^\nu(\hat{\omega}_{1\nu}{}^u{}_{u,11}
-2L\hat{A}^{EM}_\nu)
(y)}:,
\label{Q3}
&&\\
\tilde{f}(x)
=Q_4:\exp{i\int^xdy^\nu(\hat{\omega}_{2\nu}{}^u{}_{u,11}
-2L\hat{A}^{EM}_\nu)(y)}:,
\label{Q4}
&&
\end{eqnarray}
where $Q_i$, $i=1,\ldots,4$ are arbitrary constants.
$\hat{\omega}$, $\hat{A}^{EM}_\nu$ denote operators at the quantum level.
$:$ expresses a normal-ordering.
This normal ordering should be defined which can be consistent with 
(\ref{tilf1sol}), (\ref{tilf2sol}),
(\ref{f1sol}) and (\ref{f2sol}).

\section{Conclusion and discussion}

We have proposed a gravitational model coupled with
$SU(2)$ doublet complex Higgs-like fields whose couplings 
depend on the local coordinates of $M_4$.
As the results of
some on-shell conditions (the unitarity condition 
and the torsion-less condition)in the noncommutative space,
these Higgs couplings have been represented by
the Wilson loops of the connections 
and the gauge fields, (\ref{tilf1sol}), (\ref{tilf2sol}),
(\ref{f1sol}) and (\ref{f2sol}).
Moreover if we use these equations, we can {\it geometrically} present 
the conditions that the Higgs couplings become zero.
The choices of the topological properties for 
the four dimensional continuum $M_4$
are not free, but are restricted by the other conditions,
\begin{eqnarray}
|\tilde{f}|^2=-\frac{2}{v^2}\frac{\omega_{1\mu}{}^{ua}{}_{,11}}
{\omega_{1\mu}{}^{a}{}_{u,11}}
=-\frac{2}{v^2}\frac{\omega_{1\mu}{}^{ua},{}_{12}}{\omega_{1\mu}{}^a{}_{u,21}}
=\ldots
\nonumber
\end{eqnarray}
Notice that these restrictions contain the connections
which the spread over the noncommutative space $M_4\times{Z_2}$.
Therefore, we wish to stress in this paper that
if we construct the standard model on the non-commutative space
and we elevate the Higgs couplings to some geometrical objects
on the noncommutative space, we would gain
a system beyond the standard model, which could represent
the masses of leptons and quarks as geometrical things   
of noncommutative space.

Lastly, we confess what we should improve in our construction.
We have not decomposed the components of our bases containing
the local Higgs couplings as the authors have performed in \cite{CFG}.
They have described the decomposition of their one-form
as
$E^r=e^\sigma\left[D,m^r\right]$, where
$m^5=
\left(
\begin{array}{ccc}
0&-1&0\\
-1&0&0\\
0&0&0
\end{array}
\right)$,
$m^6=\left(
\begin{array}{ccc}
0&0&0\\
0&-1&0\\
0&0&0
\end{array}
\right)
$.
In other words, we have not exactly presented
the one-form of $\tilde{\Omega}^1_D({\cal A})$ 
for our algebra (\ref{ouralgebra}).
The one-form $\tilde{\Omega}^1_D({\cal A})$ 
should be represented by a zero-form times a commutator between
a zero-form and a Dirac operator.
After excluding the auxiliary part of our base given in this paper, 
a final description of 
these conditions with respect to the Higgs-like couplings
would be changed. However if we add a one-form of some algebra,
we may revive the expressions of the above Wilson loops.

We hope that our attempt will be one of the advantages to found 
a formulation beyond the standard model.

\vskip 8mm

{\sl Acknowledgments}

We would like to thank 
Professor Watamura for giving his seminar at our university
and for stimulating discussion. 
One of the authors (R.K) would also like to thank him for 
discussing about \cite{CFG} and \cite{CFF}.
After submitting version 2 of this paper to arXiv, 
R.K. had a chance to talk about 
\cite{CL}and \cite{CFG} to some members of theoretical physics group in KEK.
They commented on the renormalizability of noncommutative field theories.
R.K. would like to thank them for the valuable discussions and comments.
R.K. is supported by the Research Fellowships
of the Japan Society for the Promotion of Science.

\bigskip

\appendix

\section{Unitarity conditions}

We calculate the unitarity condition 
by using 
\begin{eqnarray}
d_{EM}{\langle
{\cal E}^A,{\cal E}^B
\rangle}_D=
-\Omega^A{}_C{\langle
{\cal E}^C,{\cal E}^B
\rangle}_D
+{\langle
{\cal E}^A,{\cal E}^C
\rangle}_D(\Omega^B{}_C)^\ast.
\end{eqnarray}
In order to calculate the unitarity condition of the bases 
and the spin connection,
we need to estimate the generalized hermitian inner product
between the above one-form bases 
as (3.30) and (3.31) in \cite{CFG},  
\begin{eqnarray}
{\langle 
{\cal E}^A,{\cal E}^B
\rangle}_D
=-\frac{1}{2}\{\gamma(A),\gamma(B)\}_+\hat{{\cal E}}^A\hat{{\cal E}}^B,
\end{eqnarray}
where $\{\gamma(A),\gamma(B)\}_+$ means the anti-commutator
between two gamma matrixes which are respectively containing ${\cal E}^A$
and ${\cal E}^B$.
$\hat{{\cal E}}^B$ means ${\cal E}^B$ removing the gamma matrix, $\gamma^B$. 
The inner products between each (\ref{Ea}), (\ref{Ebar})
(\ref{curved}) and (\ref{higgs}) 
are respectively expressed by 
\begin{eqnarray}
{\langle
{\cal E}^a,{\cal E}^b
\rangle}_D&=&\delta^{ab},
\nonumber\\
{\langle
{\cal E}^{\bar{u}},{\cal E}^{\bar{u}}
\rangle}_D&=&
\frac{v^2}{2}
\left(
\begin{array}{cccc}
1&0&0&0\\
0&0&0&0\\
0&0&1&0\\
0&0&0&0
\end{array}
\right),
\nonumber\\
{\langle
{\cal E}^{\bar{d}},{\cal E}^{\bar{d}}
\rangle}_D&=&
\frac{v^2}{2}
\left(
\begin{array}{cccc}
0&0&0&0\\
0&1&0&0\\
0&0&0&0\\
0&0&0&1
\end{array}
\right),
\nonumber\\
{\langle
{\cal E}^{\bar{u}},{\cal E}^{\bar{d}}
\rangle}_D&=&
{
\langle
{\cal E}^{\bar{d}},{\cal E}^{\bar{u}}
\rangle
}_D
=0.
\end{eqnarray}
For the bases of the curved space:
\begin{eqnarray}
{\langle
{\cal E}^\mu,{\cal E}^\nu
\rangle}_D=g^{\mu\nu}(x),~~~g^{\mu\nu}=e^\mu_ae^\nu_b\delta^{ab},
\end{eqnarray}
\begin{eqnarray}
{\langle
{\cal E}^u,{\cal E}^u
\rangle}_D&=&
{\mid{\tilde{f}(x)}\mid}^2\frac{v^2}{2}
\left(
\begin{array}{cccc}
1&0&0&0\\
0&0&0&0\\
0&0&1&0\\
0&0&0&0
\end{array}
\right),
\\
{\langle
{\cal E}^d,{\cal E}^d
\rangle}_D&=&
{\mid{f}(x)\mid}^2\frac{v^2}{2}
\left(
\begin{array}{cccc}
0&0&0&0\\
0&1&0&0\\
0&0&0&0\\
0&0&0&1
\end{array}
\right),
\\
{\langle
{\cal E}^u,{\cal E}^d
\rangle}_D&=&{\langle{{\cal E}}^d,{\cal E}^u\rangle}_D=0.
\end{eqnarray}
%
$({\bf A},{\bf B})=(a,b)$
\begin{eqnarray}
\omega_{1\mu}{}^{ab},{}_{11}+\omega_{1\mu}{}^{ba},{}_{11}&=&0,
\nonumber\\
\omega_{1\mu}{}^{ab},{}_{12}+\omega_{1\mu}{}^{ba},{}_{21}&=&0,
\nonumber\\
\omega_2{}^{ab}-\omega_1{}^{ba}=0,
\end{eqnarray}
\begin{eqnarray}
\omega_{1\mu}{}^{ab},{}_{21}+\omega_{1\mu}{}^{ba},{}_{12}&=&0,
\nonumber\\
\omega_{1\mu}{}^{ab},{}_{22}+\omega_{1\mu}{}^{ba},{}_{22}&=&0,
\nonumber\\
{\omega'}_2{}^{ab}-{\omega'}_1{}^{ab}=0,
\end{eqnarray}
\begin{eqnarray}
{\omega'}_1{}^{ab}-{\omega'}_2{}^{ba}&=&0,
\nonumber\\
\omega_{2\mu}{}^{ab},{}_{11}+\omega_{2\mu}{}^{ba},{}_{11}&=&0,
\nonumber\\
\omega_{2\mu}{}^{ab},{}_{12}+\omega_{2\mu}{}^{ba},{}_{21}=0,
\end{eqnarray}
\begin{eqnarray}
\omega_1{}^{ab}-\omega_2{}^{ba}&=&0,
\nonumber\\
\omega_{2\mu}{}^{ab},{}_{21}+\omega_{2\mu}{}^{ba},{}_{12}&=&0,
\nonumber\\
\omega_{2\mu}{}^{ab},{}_{22}+\omega_{2\mu}{}^{ba},{}_{22}=0.
\end{eqnarray}
$({\bf A},{\bf B})=(a,u)$
\begin{eqnarray}
{\mid\tilde{f}\mid}^2\frac{v^2}{2}\omega_{1\mu}{}^a{}_u,{}_{11}
+\omega_{1\mu}{}^{ua},{}_{11}&=&0,
\nonumber\\
\omega_{1\mu}{}^{ua},{}_{21}=0,~~
\omega_1{}^{ua}=0,&&
\nonumber\\
{\mid\tilde{f}\mid}^2\frac{v^2}{2}\omega_{1\mu}{}^a{}_u,{}_{21}
+\omega_{1\mu}{}^{ua},{}_{12}&=&0,
\nonumber\\
-{\mid\tilde{f}\mid}^2\frac{v^2}{2}
{\omega'}_2{}^a{}_u+
{\omega'}_1{}^{ua}&=&0,
\nonumber\\
{\omega'}_2{}^{ua}&=&0,
\nonumber\\
{\mid\tilde{f}\mid}^2\frac{v^2}{2}\omega_{2\mu}{}^a{}_u,{}_{11}
+\omega_{2\mu}{}^{ua},{}_{11}&=&0,
\nonumber\\
\omega_{2\mu}{}^{ua},{}_{21}&=&0,
\nonumber\\
-{\mid\tilde{f}\mid}^2\frac{v^2}{2}\omega_1{}^a{}_u
+\omega_2{}^{ua}&=&0,
\nonumber\\
{\mid\tilde{f}\mid}^2\frac{v^2}{2}\omega_{2\mu}{}^a{}_u,{}_{21}
+\omega_{2\mu}{}^{ua},{}_{12}&=&0,
\nonumber\\
\omega_{2\mu}{}^{ua},{}_{22}=0.
\end{eqnarray}
$({\bf A},{\bf B})=(a,d)$
\begin{eqnarray}
\omega_{1\mu}{}^{da},{}_{11}&=&0,
\nonumber\\
{\mid{f}\mid}^2\frac{v^2}{2}\omega_{1\mu}{}^a{}_d,{}_{12}
+\omega_{1\mu}{}^{6a},{}_{21}&=&0,
\nonumber\\
-{\mid{f}\mid}^2\frac{v^2}{2}\omega_2{}^a{}_d
+\omega_1{}^{da}=0,
\end{eqnarray}
\begin{eqnarray}
\omega_{1\mu}{}^{da},{}_{12}&=&0,
\nonumber\\
{\mid{f}\mid}^2\frac{v^2}{2}\omega_{1\mu}{}^a{}_d,{}_{22}
+\omega_{1\mu}{}^{da},{}_{22}&=&0,
\nonumber\\
{\omega'}_1{}^{da}=0,
\end{eqnarray}
\begin{eqnarray}
-{\mid{f}\mid}^2\frac{v^2}{2}{\omega'}_1{}^a{}_d
+{\omega'}_2{}^{da}&=&0,
\nonumber\\
\omega_{2\mu}{}^{da},{}_{11}&=&0,
\nonumber\\
{\mid{f}\mid}^2\frac{v^2}{2}\omega_{2\mu}{}^a{}_d,{}_{12}
+\omega_{2\mu}{}^{da},{}_{21}=0,
\end{eqnarray}
\begin{eqnarray}
\omega_2{}^{da}=0,~~
\omega_{2\mu}{}^{da},{}_{12}=0,&&
\nonumber\\
{\mid{f}\mid}^2\frac{v^2}{2}\omega_{2\mu}{}^a{}_d,{}_{22}
+\omega_{2\mu}{}^{da},{}_{22}=0.
\end{eqnarray}
$({\bf A},{\bf B})=(u,b)$
\begin{eqnarray}
\omega_{1\mu}{}^{ub},{}_{11}
+{\mid\tilde{f}\mid}^2\frac{v^2}{2}\omega_{1\mu}{}^b{}_u,{}_{11}&=&0,
\nonumber\\
\omega_{1\mu}{}^{ub},{}_{12}
+{\mid\tilde{f}\mid}^2\frac{v^2}{2}\omega_{1\mu}{}^b{}_u,{}_{21}&=&0,
\nonumber\\
-\omega_2{}^{ub}+{\mid\tilde{f}\mid}^2
\frac{v^2}{2}\omega_1{}^b{}_u=0,
\end{eqnarray}
\begin{eqnarray}
\omega_{1\mu}{}^{ub},{}_{21}&=&0,
\nonumber\\
\omega_{1\mu}{}^{ub},{}_{22}&=&0,
\nonumber\\
{\omega'}_2{}^{ub}=0,
\end{eqnarray}
\begin{eqnarray}
-{\omega'}_1{}^{ub}+{\mid\tilde{f}\mid}^2\frac{v^2}{2}
{\omega'}_2{}^b{}_u&=&0,
\nonumber\\
\omega_{2\mu}{}^{ub},{}_{11}-{\mid\tilde{f}\mid}^2\frac{v^2}{2}
\omega_{2\mu}{}^b{}_u,{}_{11}&=&0,
\nonumber\\
\omega_{2\mu}{}^{ub},{}_{12}+
{\mid\tilde{f}\mid}^2\frac{v^2}{2}\omega_{2\mu}{}^b{}_u,{}_{21}=0,
\end{eqnarray}
\begin{eqnarray}
\omega_1{}^{ub}&=&0,
\nonumber\\
\omega_{2\mu}{}^{ub},{}_{21}&=&0,
\nonumber\\
\omega_{2\mu}{}^{ub},{}_{22}=0.
\end{eqnarray}
$({\bf A},{\bf B})=(u,u)$
\begin{eqnarray}
i\partial^{EM}_\mu{\mid\tilde{f}\mid}^2\frac{v^2}{2}\cdot
+v^2{\mid\tilde{f}\mid}^2
\omega_{1\mu}{}^u{}_u,{}_{11}&=&0,
\label{tilf1}\\
\omega_{1\mu}{}^u{}_u,{}_{21}&=&0,
\nonumber\\
\omega_1{}^u{}_u&=&0,
\nonumber\\
{\omega'}_2{}^u{}_u&=&0,
\nonumber\\
i\partial^{EM}_\mu{\mid\tilde{f}\mid}^2\frac{v^2}{2}\cdot
+v^2{\mid\tilde{f}\mid}^2\omega_{2\mu}{}^u{}_u,{}_{11}&=&0,
\label{tilf2}\\
\omega_{2\mu}{}^u{}_u,{}_{21}=0.
\end{eqnarray}
$({\bf A},{\bf B})=(u,d)$
\begin{eqnarray}
\omega_{1\mu}{}^d{}_u,{}_{11}&=&0,
\nonumber\\
{\mid{f}\mid}^2\omega_{1\mu}{}^u{}_d,{}_{12}+
{\mid\tilde{f}\mid}^2\omega_{1\mu}{}^d{}_u,{}_{21}&=&0,
\nonumber\\
-{\mid{f}\mid}^2\omega_2{}^u{}_d+
{\mid\tilde{f}\mid}^2\omega_1{}^d{}_u&=&0,
\nonumber\\
\omega_{1\mu}{}^u{}_d,{}_{22}&=&0,
\nonumber\\
-{\mid{f}\mid}^2{\omega'}_1{}^u{}_d+
{\mid\tilde{f}\mid}^2
{\omega'}_2{}^d{}_u&=&0,
\nonumber\\
\omega_{2\mu}{}^d{}_u,{}_{11}&=&0,
\nonumber\\
{\mid{f}\mid}^2\omega_{2\mu}{}^u{}_d,{}_{12}+
{\mid\tilde{f}\mid}^2\omega_{2\mu}{}^d{}_u,{}_{21}&=&0,
\nonumber\\
\omega_{2\mu}{}^u{}_d,{}_{22}=0.
\end{eqnarray}
$({\bf A},{\bf B})=(d,b)$
\begin{eqnarray}
\omega_{1\mu}{}^{db},{}_{11}=\omega_{1\mu}{}^{db},{}_{12}
&=&\omega_2{}^{db}=0,
\nonumber\\
\omega_{1\mu}{}^{db},{}_{21}+{\mid{f}\mid}^2\frac{v^2}{2}
\omega_{1\mu}{}^b{}_d,{}_{12}&=&0,
\nonumber\\
\omega_{1\mu}{}^{db},{}_{22}+{\mid{f}\mid}^2\frac{v^2}{2}
\omega_{1\mu}{}^b{}_d,{}_{22}&=&0,
\nonumber\\
-\acute{\omega}_2{}^{db}+{\mid{f}\mid}^2
\frac{v^2}{2}{\omega'}_{1}{}^b{}_d&=&0,
\nonumber\\
{\omega'}_1{}^{db}=\omega_{2\mu}^{db},{}_{11}
=\omega_{2\mu}{}^{db},{}_{12}&=&0,
\nonumber\\
-\omega_1{}^{db}+{\mid{f}\mid}^2\frac{v^2}{2}\omega_2{}^b{}_d&=&0,
\nonumber\\
\omega_{2\mu}{}^{db},{}_{21}+{\mid{f}\mid}^2\frac{v^2}{2}
\omega_{2\mu}{}^b{}_d,{}_{12}&=&0,
\nonumber\\
\omega_{2\mu}{}^{db},{}_{22}+
{\mid{f}\mid}^2\frac{v^2}{2}
\omega_{2\mu}{}^b{}_d,{}_{22}=0.
\end{eqnarray}
$({\bf A},{\bf B})=(d,u)$
\begin{eqnarray}
\omega_{1\mu}{}^d{}_u,{}_{11}&=&0,
\nonumber\\
{\mid\tilde{f}\mid}^2\omega_{1\mu}{}^d{}_u,{}_{21}
+{\mid{f}\mid}^2
\omega_{1\mu}{}^u{}_d,{}_{12}&=&0,
\nonumber\\
\omega_{1\mu}{}^u{}_d,{}_{22}&=&0,
\nonumber\\
-{\mid\tilde{f}\mid}^2{\omega'}_2{}^d{}_u
+{\mid{f}\mid}^2{\omega'}_1{}^u{}_d&=&0,
\nonumber\\
\omega_{2\mu}{}^d{}_u,{}_{11}&=&0,
\nonumber\\
-{\mid\tilde{f}\mid}\omega_1{}^d{}_u+{\mid{f}\mid}^2\omega_2{}^u{}_d&=&0,
\nonumber\\
{\mid\tilde{f}\mid}^2\omega_{2\mu}{}^d{}_u,{}_{21}
+{\mid{f}\mid}^2\omega_{2\mu}{}^u{}_d,{}_{12}&=&0,
\nonumber\\
\omega_{2\mu}{}^u{}_d,{}_{22}=0.
\end{eqnarray}
$({\bf A},{\bf B})=(d,d)$
\begin{eqnarray}
\omega_{1\mu}{}^d{}_d,{}_{12}=\omega_2{}^d{}_d&=&0,
\nonumber\\
i\partial^{EM}_\mu{\mid{f}\mid}^2\frac{v^2}{2}\cdot
+{\mid{f}\mid}^2v^2\omega_{1\mu}{}^d{}_d,{}_{22}&=&0,
\label{f1}\\
{\omega'}_1{}^d{}_d&=&0,
\nonumber\\
\omega_{2\mu}{}^d{}_d,{}_{12}&=&0,
\nonumber\\
i\partial^{EM}_\mu
{\mid{f}\mid}^2\frac{v^2}{2}\cdot+
{\mid{f}\mid}^2v^2
\omega_{2\mu}{}^d{}_d,{}_{22}=0.
\label{f2}
\end{eqnarray}
 From (\ref{tilf1}) and (\ref{tilf2}), we obtain
\begin{eqnarray}
{\mid\tilde{f}\mid}^2(x)
=C_1
\exp{i\int^xdy^\nu
(2\omega_{1\nu}{}^u{}_u,{}_{11}-LA^{EM}_\nu)}&(y),&
\label{tilf1sol}\\
{\mid\tilde{f}\mid}^2(x)
=C_2
\exp{i\int^xdy^\nu
(2\omega_{2\nu}{}^u{}_u,{}_{11}-LA^{EM}_\nu)}(y).
\label{tilf2sol}
\end{eqnarray}
 From (\ref{f1}) and (\ref{f2}),
we obtain
\begin{eqnarray}
{\mid{f}\mid}^2(x)
=C_3
\exp{i\int^xdy^\nu
(2\omega_{1\nu}{}^d{}_d,{}_{22}-LA^{EM}_\nu)}&(y),&
\label{f1sol}\\
{\mid{f}\mid}^2(x)
=C_4
\exp{i\int^xdy^\nu
(2\omega_{2\nu}{}^d{}_d,{}_{22}-LA^{EM}_\nu})(y).
\label{f2sol}
\end{eqnarray}
%

\section{Components of torsion}

We have calculated the torsion $T^A=dE^A+\Omega^A{}_B{E}^B$ where
$dE^A=\left[\hat{D}_{EM}~,~E^A\right]$ and (\ref{higgs})
\footnote{As we described in the discussion, we did not eliminate
the auxiliary part which has been defined by (\ref{aux1}) and 
(\ref{aux2}) from (\ref{curved})}.
The each component of $T^A$ is given by
\begin{eqnarray}
T^a_{11}=\gamma^\mu\gamma^b\omega_{1\mu}{}^a{}_{b,11}
+i\gamma^\mu\partial_\mu\gamma^a\cdot
+\left[\gamma^\mu,\gamma^a\right]_{-}i\partial^{EM}_\mu
&&\nonumber\\
T^a_{12}=\gamma^\mu\gamma^b\omega_{1\mu}{}^a{}_{b,12}
-{\bf 1}_{4\times{4}}\exp{(-\sigma)}f^2\frac{v^2}{2}
\omega_2{}^a{}_d
&&\nonumber\\
T^a_{13}=\gamma^\mu\gamma^5\frac{v}{\surd{2}}\omega_{1\mu}{}^a{}_{u,11}
\tilde{f}^\ast+\frac{v}{\surd2}\left[\gamma^5,\gamma^a\right]_{-}
f^\ast
&&\nonumber\\
T^a_{14}=\gamma^\mu\gamma^5\frac{v}{\surd{2}}f^\ast
\omega_{1\mu}{}^a{}_{d,12}+\frac{v}{\surd{2}}f\gamma^5\gamma^b\exp{(-\sigma)}
\omega_2{}^a{}_b
&&\nonumber\\
T^a_{21}=\gamma^\mu\gamma^b\omega_{1\mu}{}^a{}_{b,21}
-{\bf 1}_{4\times{4}}\exp{(-\sigma)}\frac{v^2}{2}\tilde{f}^2
{\omega'}_2{}^a{}_u
&&\nonumber\\
T^a_{22}=\gamma^\mu\gamma^b\omega_{1\mu}{}^a{}_{b,22}
+\gamma^\mu{i}\partial_\mu\gamma^a\cdot
+\left[\gamma^\mu,\gamma^a\right]_{-}i\partial^{EM}_\mu
&&\nonumber\\
T^a_{23}=\gamma^5\gamma^b\exp{(-\sigma)}\tilde{f}
\frac{v}{\surd{2}}{\omega'}_2{}^a{}_b
+\gamma^\mu\gamma^5\omega_{1\mu}{}^a{}_{u,21}\frac{v}{\surd{2}}
\tilde{f}^\ast
&&\nonumber\\
T^a_{24}=
\gamma^\mu\gamma^5\frac{v}{\surd2}f^\ast\omega_{1\mu}{}^a{}_{d,22}
+\frac{v}{\surd2}\tilde{f}^\ast\left[\gamma^5,\gamma^a\right]_{-}
&&\nonumber\\
T^a_{31}=
-\gamma^\mu\gamma^5\tilde{f}\frac{v}{\surd2}\omega_{2\mu}{}^a{}_{u,11}
+\frac{v}{\surd2}\left[\gamma^5,\gamma^a\right]_{-}f
&&\nonumber\\
T^a_{32}=
{\omega'}_1{}^a{}_b\tilde{f}^\ast\frac{v}{\surd2}
\gamma^5\gamma^b\exp{(-\sigma)}-\frac{v}{\surd2}f\gamma^\mu\gamma^5
\omega_{2\mu}{}^a{}_{d,12}
&&\nonumber\\
T^a_{33}=i\gamma^\mu\partial_\mu\gamma^a\cdot
+\left[\gamma^\mu,\gamma^a\right]_{-}i\partial^{EM}_\mu
+\gamma^\mu\gamma^b\omega_{2\mu}{}^a{}_{b,11}
&&\nonumber\\
T^a_{34}=\gamma^\mu\gamma^b\omega_{2\mu}{}^a{}_{b,12}
+{\bf 1}_{4\times{4}}\exp{(-\sigma)}\frac{v^2}{2}
f^\ast\tilde{f}^\ast{\omega'}_1{}^a{}_d
&&\nonumber\\
T^a_{41}=\omega_1{}^a{}_bf^\ast\frac{v}{\surd2}
\gamma^5\gamma^b\exp{(-\sigma)}-\frac{v}{\surd2}\gamma^\mu\gamma^5
\tilde{f}\omega_{2\mu}{}^a{}_{u,21}
&&\nonumber\\
T^a_{42}=
\frac{v}{\surd2}\left[\gamma^5,\gamma^a\right]_{-}\tilde{f}
-\frac{v}{\surd2}f\gamma^\mu\gamma^5\omega_{2\mu}{}^a{}_{d,22}
&&\nonumber\\
T^a_{43}=
\gamma^\mu\gamma^b\omega_{2\mu}{}^a{}_{b,21}
+{\bf 1}_{4\times{4}}\exp{(-\sigma)}\tilde{f}^\ast{f}^\ast
\frac{v^2}{2}\omega_1{}^a{}_u
&&\nonumber\\
T^a_{44}=\gamma^\mu\gamma^b\omega_{2\mu}{}^a{}_{b,22}
+\gamma^\mu{i}\partial_\mu\gamma^a\cdot
+\left[\gamma^\mu,\gamma^a\right]_{-}i\partial^{EM}_{\mu}
&&
\end{eqnarray}
%
\begin{eqnarray}
T^d_{11}=\gamma^\mu\gamma^b\omega_{1\mu}{}^d{}_{b,11}&&
\nonumber\\
T^d_{12}=\gamma^\mu\gamma^b\omega_{1\mu}{}^d{}_{b,12}
-{\bf 1}_{4\times{4}}\exp{(-\sigma)}f^2\frac{v^2}{2}
\omega_2{}^d{}_d
&&
\nonumber\\
T^d_{13}=\gamma^\mu\gamma^5\frac{v}{\surd2}\omega_{1\mu}{}^d{}_{u,11}
\tilde{f}^\ast&&
\nonumber\\
T^d_{14}=\gamma^\mu\gamma^5\frac{v}{\surd2}f^\ast\omega_{1\mu}{}^d{}_{d,12}
&&
\nonumber\\
T^d_{21}=\gamma^\mu\gamma^b\omega_{1\mu}{}^d{}_{b,21}
-{\bf 1}_{4\times{4}}\exp{(-\sigma)}\frac{v^2}{2}
{\omega'}_2{}^d{}_u&&
\nonumber\\
T^d_{22}=-\frac{v^2}{2}(\tilde{f}^\ast{f}+f^\ast\tilde{f})
{\bf 1}_{4\times{4}}
&&
\nonumber\\
T^d_{23}=\gamma^5\gamma^b\exp{(-\sigma)}\tilde{f}\frac{v}{\surd2}
{\omega'}_2{}^d{}_b
+\gamma^\mu\gamma^5\omega_{1\mu}{}^d{}_{u,21}
\frac{v}{\surd2}\tilde{f}^\ast
&&
\nonumber\\
T^d_{24}=\gamma^\mu\gamma^5\frac{v}{\surd2}f^\ast\omega_{1\mu}{}^d{}_{d,22}
+\frac{v}{\surd2}(i\gamma^\mu\gamma^5\partial_\mu{f}^\ast\cdot
-\left[\gamma^5,\gamma^\mu\right]_{-}f^\ast{i}\partial_{EM\mu})
&&
\nonumber\\
T^d_{31}=-\gamma^\mu\gamma^5\tilde{f}\frac{v}{\surd2}
\omega_{2\mu}{}^d{}_{u,11}&&
\nonumber\\
T^d_{32}={\omega'}_1{}^d{}_b\tilde{f}^\ast
\frac{v}{\surd2}\gamma^5\gamma^b\exp{(-\sigma)}
-\frac{v}{\surd2}f\gamma^\mu\gamma^5
\omega_{2\mu}{}^d{}_{d,12}&&
\nonumber\\
T^d_{33}=\gamma^\mu\gamma^b\omega_{2\mu}{}^d{}_{b,11}
&&
\nonumber\\
T^d_{34}=\gamma^\mu\gamma^b\omega_{2\mu}{}^d{}_{b,12}
+{\bf 1}_{4\times{4}}\exp{(-\sigma)}
\frac{v^2}{2}{f}^\ast\tilde{f}^\ast
\acute{\omega}_1{}^d{}_d
&&
\nonumber\\
T^d_{41}=\omega_1{}^d{}_b{f}^\ast\frac{v}{\surd2}\gamma^5\gamma^b
\exp{(-\sigma)}-\frac{v}{\surd2}\gamma^\mu\gamma^5\tilde{f}
\omega_{2\mu}{}^d{}_{u,21}
&&
\nonumber\\
T^d_{42}=-\frac{v}{\surd2}
(i\gamma^\mu\gamma^5\partial_\mu{f}\cdot
-\left[\gamma^5,\gamma^\mu\right]_{-}
f{i}\partial_{EM\mu})
-\frac{v}{\surd2}f\gamma^\mu\gamma^5\omega_{2\mu}{}^d{}_{d,22}
&&
\nonumber\\
T^d_{43}=\gamma^\mu\gamma^b\omega_{2\mu}{}^d{}_{b,21}
+{\bf 1}_{4\times{4}}\exp{(-\sigma)}\tilde{f}^\ast
f^\ast\frac{v^2}{2}\omega_1{}^d{}_u
&&
\nonumber\\
T^d_{44}=\gamma^\mu\gamma^b\omega_{2\mu}{}^d{}_{b,22}
+\frac{v^2}{2}(\tilde{f}f^\ast+f\tilde{f}^\ast){\bf 1}_{4\times{4}}
&&
\end{eqnarray}
\begin{eqnarray}
T^u_{11}=-\frac{v^2}{2}{\bf 1}_{4\times{4}}(f^\ast\tilde{f}
+\tilde{f}^\ast{f})+\gamma^\mu\gamma^b\omega_{1\mu}{}^u{}_{b,11}
&&
\nonumber\\
T^u_{12}=\gamma^\mu\gamma^b\omega_{1\mu}{}^u{}_{b,12}-
{\bf 1}_{4\times{4}}\exp{(-\sigma)}f^2
\frac{v^2}{2}\omega_2{}^u{}_d
&&
\nonumber\\
T^u_{13}=\gamma^\mu\gamma^5\frac{v}{\surd2}\omega_{1\mu}{}^u{}_{u,11}
\tilde{f}^\ast
+i\frac{v}{\surd2}\gamma^\mu\gamma^5\partial_\mu\tilde{f}^\ast\cdot
+i\frac{v}{\surd2}\tilde{f}^\ast
\left[\gamma^\mu,\gamma^5\right]_{-}
\partial_{EM\mu}
&&
\nonumber\\
T^u_{14}=\gamma^\mu\gamma^5\frac{v}{\surd2}f^\ast
\omega_{1\mu}{}^u{}_{d,12}
+f\frac{v}{\surd2}\gamma^5\gamma^b\exp{(-\sigma)}\omega_2{}^u{}_b
&&
\nonumber\\
T^u_{21}=\gamma^\mu\gamma^b\omega_{1\mu}{}^u{}_{b,21}
-{\bf 1}_{4\times{4}}\exp{(-\sigma)}\frac{v^2}{2}
\tilde{f}^2{\omega'}_2{}^u{}_u
&&
\nonumber\\
T^u_{22}=\gamma^\mu\gamma^b\omega_{1\mu}{}^u{}_{b,22}
&&
\nonumber\\
T^u_{23}=\gamma^5\gamma^b\exp{(-\sigma)}\tilde{f}\frac{v}{\surd2}
{\omega'}_2{}^u{}_b
+\gamma^\mu\gamma^5\omega_{1\mu}{}^u{}_{u,21}\frac{v}{\surd2}\tilde{f}^\ast
&&
\nonumber\\
T^u_{24}=\gamma^\mu\gamma^5\frac{v}{\surd2}f^\ast\omega_{1\mu}{}^u{}_{d,22}
&&
\nonumber\\
T^u_{31}=
-i\frac{v}{\surd2}\gamma^\mu\gamma^5
\partial_\mu\tilde{f}\cdot-\frac{v}{\surd2}
\tilde{f}\left[\gamma^\mu,\gamma^5\right]_{-}i\partial_{EM\mu}
-\gamma^\mu\gamma^5\tilde{f}\frac{v}{\surd2}\omega_{2\mu}{}^u{}_{u,11}
&&
\nonumber\\
T^u_{32}={\omega'}_1{}^u{}_b\tilde{f}^\ast\frac{v}{\surd2}
\gamma^5\gamma^b\exp{(-\sigma)}-\frac{v}{\surd2}f\gamma^\mu\gamma^5
\omega_{2\mu}{}^u{}_{d,12}
&&
\nonumber\\
T^u_{33}=\gamma^\mu\gamma^b\omega_{2\mu}{}^u{}_{b,11}
+\frac{v^2}{2}
{\bf 1}_{4\times{4}}(f\tilde{f}^\ast+\tilde{f}f^\ast)
&&
\nonumber\\
T^u_{34}=\gamma^\mu\gamma^b\omega_{2\mu}{}^u{}_{b,12}
+{\bf 1}_{4\times{4}}\exp{(-\sigma)}\frac{v^2}{2}
f^\ast\tilde{f}^\ast{\omega'}_1{}^u{}_d
&&
\nonumber\\
T^u_{41}=\omega_1{}^u{}_b{f}^\ast\frac{v}{\surd2}\gamma^5\gamma^b
\exp{(-\sigma)}-\frac{v}{\surd2}\gamma^\mu\gamma^5\tilde{f}
\omega_{2\mu}{}^u{}_{u,21}
&&
\nonumber\\
T^u_{42}=-\frac{v}{\surd2}f\gamma^\mu\gamma^5\omega_{2\mu}{}^u{}_{d,22}
&&
\nonumber\\
T^u_{43}=\gamma^\mu\gamma^b\omega_{2\mu}{}^u{}_{b,21}
+{\bf 1}_{4\times{4}}\exp{(-\sigma)}\tilde{f}^\ast{f}^\ast
\frac{v^2}{2}\omega_2{}^u{}_u
&&
\nonumber\\
T^u_{44}=\gamma^\mu\gamma^b\omega_{2\mu}{}^u{}_{b,22}
&&
\end{eqnarray}
In section three we have imposed the torsion-less
condition for a part of these components.


\end{document}